\documentclass[12pt,eqsecnum,preprint]{aastex}

\slugcomment{in prep.}

\shorttitle{Radio Constraints on Activity in Young Brown Dwarfs}
\shortauthors{R. A. Osten \& R. Jayawardhana }

\begin{document}

\title{ Radio Constraints on Activity in Young Brown Dwarfs
}
\author{Rachel A. Osten\altaffilmark{1}}
\affil{Department of Astronomy, University of Maryland, College Park, MD 20742-2421, U.S.A. 
\\Electronic Mail: rosten@astro.umd.edu}
\altaffiltext{1}{Hubble Fellow}

\author{Ray Jayawardhana}
\affil{Department of Astronomy \& Astrophysics, University of Toronto, Toronto, 
ON M5S 3H8, CANADA \\Electronic Mail: rayjay@astro.utoronto.ca}

\begin{abstract}
We report on searches for radio emission from three of the nearest known young brown 
dwarfs using the Very Large Array. We have obtained sensitive upper limits on 3.6cm 
emission from 2MASSW J1207334-393254, TWA~5B and SSSPM J1102-3431, all of which are 
likely members of the $\sim$8-Myr-old TW Hydrae association. We derive constraints 
on the magnetic field strength and the number density of accelerated electrons, 
under the assumption that young brown dwarf atmospheres are able to produce 
gyrosynchrotron emission, as seems to be indicated in older brown dwarfs.
For the young brown dwarf TWA~5B, 
the ratio of its detected X-ray luminosity to the upper limit on radio luminosity
places it within the expected range for young stars and older, active stars. 
Thus, its behavior is anomalous compared to older brown dwarfs, in which radio
luminosity is substantially enhanced over the expected relationship.
Our observations deepen the conundrum of magnetic activity in brown dwarfs, 
and suggest that a factor other than age is more important for determining radio 
emission in cool substellar objects.
\end{abstract}


\section{Introduction }
Several lines of evidence point to young brown dwarfs undergoing a T Tauri phase
similar to low-mass pre-main sequence stars: near- and mid-infrared excesses 
consistent with dusty disks \citep[e.g.,][]{natta2002,jetal2003}, 
broad, asymmetric H$\alpha$ emission associated with accretion \citep[e.g.,][]{jmb2003}
irregular photometric and spectroscopic variability 
(Scholz \& Eisl\"offel 2004; Scholz \& Jayawardhana 2005), and forbidden emission 
lines believed to originate in outflows \citep{fc2001,bnj2004}. 
The H$\alpha$ activity seen in 
(non-accreting) young brown dwarfs is comparable to saturated activity levels 
in field M dwarfs with similar spectral type and rotation rates \citep{jmb2002}.

The X-ray properties of young brown dwarfs appear to be similar to young low-mass stars
at $\sim$ 1 Myr in age, according to a recent report by \citet{preibisch2005}.
Another study by \citet{stelzer2005} suggests that the X-ray emission of brown dwarfs with the
same effective temperature have similar $L_{X}/L_{bol}$ for a range of ages, from young BDs
to field objects.  Thus young brown dwarfs appear to share similarities in this magnetic 
activity signature both with objects of the same age but higher mass and with 
objects of the same temperature but different age.

In T Tauri stars, magnetic activity (radio and X-ray) is enhanced compared to field 
stars of Gyr ages. The ratio of radio and X-ray luminosities of a heterogeneous 
mixture of active stars, including T Tauri stars,
has a constant level, L$_{R}\sim L_{X}/$10$^{15.5}$ Hz$^{-1}$ \citep{gb1993}.
Radio emission from wTTs ranges in luminosity
from 10$^{15}$--10$^{18}$ erg s$^{-1}$ Hz$^{-1}$, and is generally attributed to a nonthermal
mechanism 
\citep[gyrosynchrotron emission by analogy with the Sun and active stars;][]{white1992}.
CTTs are not detected in the radio as frequently as WTTS; the few that are show emission
consistent with thermal bremsstrahlung from ionized winds \citep{martin1996}.

For field brown dwarfs of Gyr ages, magnetic activity detections 
(primarily radio, X-ray, H$\alpha$)
are more sporadic, and do not follow the activity ratios of active stars.
The underlying physical mechanism giving rise to the 
quiescent radio emission, at least in one case \citep{osten2006}, 
 appears to be similar to that seen in active stars -- namely,
gyrosynchrotron emission from a population of mildly relativistic electrons.  
This is in line with the X-ray studies that demonstrate a similarity of 
emission properties and mechanisms. 

In this Letter, we investigate whether the radio emission properties seen in 
T Tauri stars extend to young brown dwarfs, and whether radio emission from brown 
dwarfs is enhanced in their youth. We focus on the nearby TW~Hydrae association, 
which has several known brown dwarf members.  Its age is sufficiently
young ($\sim$8Myr) that signatures of accretion and magnetic activity are still observable. 
Hence, we expect to see analagous radio properties to T Tauri stars with an without 
disks.

\section{Targets}
The vast majority of known young brown dwarfs reside in star-forming regions or clusters 
at distances $\geq$ 140 pc, making it extremely difficult, if not impossible, to detect 
their radio emission with current facilities. However, a few young brown dwarfs have 
been identified recently in stellar associations located much closer to the Sun. Here 
we focus on three of the four known sub-stellar members of the TW Hydrae association at 
$\sim$50 pc. Given their relative proximity, these objects offer the best chance of 
constraining radio emisson from young brown dwarfs. 

\subsection{2MASSW J1207334-393254}

\citet{gizis2002} identified 2MASSW J1207334-393254 (hereafter 2M1207) as a likely member 
of the TW Hydrae association, based on its strong H$\alpha$ emission, low surface 
gravity and space motion. The case for its youth and membership was strengthened by 
follow up high-resolution optical spectroscopy \citep{mjbn2003}.  Intriguingly, it shows
strong emission both in the Hydrogen Balmer series (H$\alpha$ to H$\epsilon$) and
in He~I (4471, 5876, 6678, and 7065 \AA), compared with other young brown
dwarfs of similar spectral type.  The H$\alpha$ line is also relatively broad (1200 km/s)
and asymmetric.  These characteristics suggest that 2M1207 is a (weak) accretor or
an exceptionally active object. Recently, \citet{sjb2005} have 
found emission line variations on timescales of both several weeks and several hours, 
and interpret them as evidence of non-steady accretion. There is evidence for a 
dusty disk around this object, in the form of mid-infrared excess 
\citep{jetal2003,sterzik2004}.
Forbidden [ O I ] emission seen in 2M1207 at least 
in some epochs is normally associated with winds or outflows in T Tauri stars 
\citep{mjb2005}. An X-ray observation by \citet{gizisbharat} 
resulted in an upper limit to the X-ray luminosity of 1.2$\times$10$^{26}$ erg s$^{-1}$.

2M1207 has a planetary mass companion located 778 mas away \citep{chauvin2004}.
Using the moving cluster method, \citet{mamajek2005} has recently revised 2M1207's 
distance estimate from 70 pc to 53$\pm$6 pc. We adopt this revised distance
in our calculations. Given the relatively high mass of the newly found companion (20\% of
the primary), its wide separation, and the fact that brown dwarf disks appear to be 
very low mass \citep[e.g.][]{klein2003}, this system may have formed as a binary 
brown dwarf, rather than the companion forming out of a `protoplanetary' disk.
Especially in that case, the companion itself may be chromospherically active
and may even have its own wind.

\subsection{TWA~5B}
TWA~5B is a companion to the M1.5 T Tauri star TWA~5, discovered by \citet{lowrance1999}, 
at an apparent separation of 2''. It was detected by \citet{tsuboi2003} in X-rays, 
with an X-ray luminosity of 4$\times$10$^{27}$ erg s$^{-1}$, assuming a distance of 50 pc.  
\citet{mamajek2005} used the moving cluster method to estimate a distance to TWA5 
of 44$\pm$4pc, which we adopt.

\subsection{SSSPM J1102-3431}
This object was identified as a probable sub-stellar member of the TW Hydrae association by 
\citet{scholz2005}. Given its apparent proximity to the T Tauri star TW Hya, these authors 
noted that the two could form a wide binary. \citet{sjb2005} presented a high resolution 
optical spectrum of SSSPM J1102-3431 exhbiting narrow and symmetric emission lines, which 
suggest that only weak accretion is taking place, if at all. 
\citet{mamajek2005} used the moving cluster method to estimate a distance of 43$\pm$7 pc 
to this object. 

\section{Observations}
The Very Large Array \footnote{The National Radio Astronomy Observatory is a facility of the National Science Foundation operated under cooperative agreement by Associated Universities, Inc.} 
observed three of the four known TW~Hya brown dwarfs 
at 3.6 cm. 
Table~\ref{tbl:log} displays the log of the observations.
We made use of an archival observation of TWA~5 in A array (from program AC605)
which had sufficient spatial resolution to isolate the brown dwarf companion
from the T Tauri star, although not enough integration time to be very sensitive.
The data were reduced and calibrated using AIPS software (version 31 Dec03).
The observations of 2M1207 were collected in A array on 7 different days 
from 2005 January 11--19 to accumulate the total on-source time. 
Each dataset was calibrated and imaged separately and no source was detected.  
Finally, the visibility datasets were combined to facilitate weak point source detection.

\section{Results}
No sources were detected at the location of 2M1207 or its planetary mass companion,
nor in maps at the positions of SSSPM J1102-3431 or TWA~5B.
A preliminary analysis of the dataset on TWA~5B, presented in

We determined the rms variations in the images using natural weighting. 
We report the 3 $\sigma$ noise estimate in Table~\ref{tbl:log}
as our 99\% confidence limit on the 3.6 cm flux
densities.

\section{Constraints on Coronal Properties}
If young brown dwarfs exhibit accretion signatures and coronal activity (X-ray) signatures
with characteristics similar to T Tauri stars, does this similarity extend to radio emission
as well?
The situation is somewhat more complicated, as 
radio emission from T Tauri stars can be either thermal bremsstrahlung or gyrosynchrotron
radiation: classical T Tauri stars tend to show thermal radio emission, whereas
weak-lined T Tauri stars generally display nonthermal emission from 
electrons accelerated in the presence of a magnetic field.  
\citet{andre1987} suggested that stars with disks may also be emitting non-thermal 
radio radiation, but it gets absorbed by high density stellar winds.  



\begin{figure}[htbp]
\includegraphics[scale=0.5]{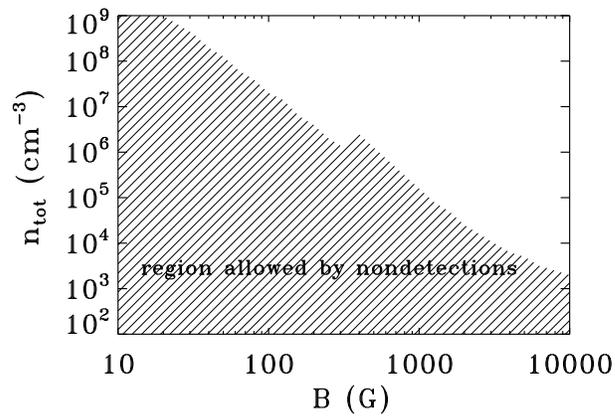}
\caption{Region of space constrained by the nondetection of radio flux from 2M1207.
We assume that the emission mechanism is gyrosynchrotron, from a relatively hard population
of nonthermal electrons within a dipole magnetic field configuration, and allow the 
electron density to have no radial dependence (i.e. constant with radius).  The shaded section 
indicates
locations 
consistent with the 3$\sigma$ upper limit of 29$\mu$Jy on 2M1207 at a distance of 53 pc.
 \label{fig:constraints}
}
\end{figure}

If we assume that a magnetic field accelerates electrons, we can use the analytic 
models of \citet{whiteetal1989} to investigate possible combinations of parameters.  For
simplicity, we assume there is a global dipole field, and the nonthermal electrons have no
radial falloff with distance.  We take the index of the power-law distribution of electrons
to be $\delta=3$, in line with inferences from active early- and late-type M stars that 
suggest such a hard distribution \citep{whiteetal1989,osten2006}, and constrain the dipole 
size to be the object's radius, $\sim$0.5 R$_{\odot}$ \citep{mjb2004}.
With these simplifying assumptions, the free parameters are the base magnetic field strength
and total number density of nonthermal electrons above a cut-off energy.  Figure~\ref{fig:constraints} displays
the locus of points in magnetic field-electron density space compatible with our upper limit
at 3.6 cm.  If young brown dwarfs harbor kilogauss magnetic fields, as \citet{sj2005} 
suggests, then this constrains the total number density of accelerated electrons to be 
less than  $n_{\rm tot}\sim$ 10$^{5}$cm$^{-3}$ or so.


\section{Discussion}

\citet{claussen2001} investigated centimeter-wavelength radio emission from
T Tauri stars in the TW Hydrae association, detecting 5 out of 11 with luminosities  
in the range of 
5.6--320$\times$10$^{14}$ erg s$^{-1}$ Hz$^{-1}$, assuming a distance of 50 pc.
The radio spectrum of TW~Hya is consistent with partially optically thick thermal free-free emission.
The upper limits on TWA brown dwarf radio luminosities 
in Table~\ref{tbl:log} are factors of 3--300 times less than those detected for the T Tauri stars
at similar distances.  

The steady luminosities of field brown dwarfs and very low mass stars detected at radio 
wavelengths, with the same spectral types as the young BD discussed here,  
span $\sim$4$\times$10$^{12}$--3$\times$10$^{13}$
erg s$^{-1}$ Hz$^{-1}$ \citep{bergeretal2001,osten2006}.
Our upper limits suggest that enhancements of radio emission by more than a factor 
of 3--25 do not occur in these young brown dwarfs compared with field brown dwarfs. 
This assumes that the underlying emission mechanism is the same, and that age is the primary
controller of radio emission.  Studies of slower rotating, young T Tauri stars 
\citep{mamajek1999} suggest that rotation may be a more important factor than age in 
producing detectable radio emission. 
This suggestion is bolstered by \citet{berger2002}, who pointed out that field brown dwarfs 
exhibiting radio emission were systematically faster rotaters than undetected objects.  
Two of the three objects investigated here have relatively low v$\sin i$, 
$\leq$ 26 km s$^{-1}$, whereas four of the six field low-mass stars/brown dwarfs
discussed in \citet{berger2002} and \citet{bp2005} have v$\sin i\ge$25 km s$^{-1}$.  
This result is only suggestive; a larger sample is needed for confirmation. 

The detection of X-ray emission from TWA~5B is indicative, at least in this object, of an
enhancement of magnetic activity with youth.  Its X-ray luminosity of 4$\times$10$^{27}$
erg s$^{-1}$ \citep{tsuboi2003} is a factor of 20--200 times larger than that of field low mass stars
of similar spectral type \citep[LHS 2065 and VB 10,][respectively]{sl02,f03}.
We would expect, by analogy with T Tauri stars, that there
should be radio emission from TWA~5B associated with magnetic activity, and if the 
L$_{X}$--L$_{R}$ relation holds
for TWA~5B as well, we would expect L$_{R}\sim$1.2$\times$10$^{12}$ erg s$^{-1}$ Hz$^{-1}$.  
This is about two orders of magnitude below our 3$\sigma$ upper limit, so we cannot rule out
the possibility that plasma heating and particle acceleration are occuring in the same ratio
in TWA~5B as for active stars.
However, the ratio of radio to X-ray luminosities in field brown dwarfs
and very low mass stars diverges from that seen in active stars \citep{bergeretal2001,berger2002}, 
but it is in the 
opposite sense as seen here: the radio emission exceeds the L$_{X}$--L$_{R}$ relation 
by more than
four orders of magnitude.
If TWA~5B were to behave like the field brown dwarfs detected at radio wavelengths, 
we would expect its radio emission to be 
1.2$\times$10$^{16}$ erg s$^{-1}$ Hz$^{-1}$, or more than two orders of magnitude
higher than our upper limit.
Thus it appears that radio emission from young brown dwarfs displays a
disconnect with the behavior demonstrated by field brown dwarfs of Gyr age.


Another phenomenon seen in pre-main sequence stars, very low mass stars and
brown dwarfs is variability, interpreted as magnetic reconnection flares when
varying radio, X-ray, and UV emissions are seen \citep{bergeretal2001,sl02,hjk2003}.
Particularly in X-rays, flaring signatures are seen in the absence of evidence of
steady emission \citep{rutledge2000}.
Our long observation of 2M1207 encompassed 8.4 hours; if we assume
that a typical flare time scale is $\sim$30 minutes, then our lack of radio detection 
implies a flare duty cycle less than 6\%, which is still consistent with flare
duty cycles from radio detected field brown dwarfs \citep{osten2006}.

\section{Conclusions}
This study presents the first attempt to extend studies of activity in young brown dwarfs
to radio wavelengths.  The results suggest that the anomalous behavior of the radio
emission in field brown dwarfs is not primarily a function of age, and may depend on another
parameter; rotation has been suggested as a key on two different fronts to controlling
the radio emission properties of both young stars and field brown dwarfs.
For the near future, further radio studies are limited to the nearest young 
groups due to sensitivity constraints.  
The {\it Expanded VLA} will be instrumental in carrying out these observations, with
improvements of up to a factor of 5 for $\nu \leq$ 10 GHz, which will allow similar
luminosity constraints on 
clusters up to $\sim$ 100 pc, probing a wider array of star/brown dwarf formation characteristics.
Larger samples are needed to explore the dependence of radio emission-induced magnetic activity 
on temperature, mass, and age in the sub-stellar regime.


This paper represents the results of VLA projects AO190, AO192, and AO194.

\acknowledgments Support for this work was provided by NASA through Hubble Fellowship 
grant \# HF-01189.01
awarded by the Space Telescope Science Institute, which is operated by the Association
of Universities for Research in Astronomy, Inc. for NASA, under contract NAS5-26555.
RAO is grateful to M. Claussen for providing results prior to publication. RJ acknowledges
support from an NSERC grant. 


\begin{deluxetable}{lllllll}
\tablewidth{0pt}
\tablecolumns{7}
\tablecaption{TW HYa Brown Dwarf Properties \label{tbl:prop}}
\tablehead{\colhead{Source} & \colhead{sp. type} &\colhead{distance\tablenotemark{a}} & \colhead{vsini} & \colhead{L$_{X}$}
&\colhead{H$\alpha$ 10\% width} &\colhead{Refs\tablenotemark{b}}\\
\colhead{} & \colhead{} &\colhead{(pc)} & \colhead{(km s$^{-1}$)} & \colhead{(10$^{27}$ erg s$^{-1}$)} & 
\colhead{(km s$^{-1}$)} & \colhead{} }
\startdata
2M1207 & M8.0 & 53$\pm$3 & 13.0 &$<$0.12   & 200--300 &MJBN03,GB04,SJB05   \\
TWA 5B & M8.5 &  44$\pm$4 & 26.0 &4   & 160  & MJBN03,T03,MJB05\\
SSSPM 1102 & M8.5 &43$\pm$7  & ---   &---   &190  & SJB05\\
\enddata
\tablenotetext{a}{Distances taken from \citet{mamajek2005}.}
\tablenotetext{b}{MJBN03=\citet{mjbn2003}, GB04=\citet{gizisbharat},SJB05=\citet{sjb2005},
T03=\citet{tsuboi2003},MJB05=\citet{mjb2005}}
\end{deluxetable}

\begin{deluxetable}{lllllll}
\tablewidth{0pt}
\tablecolumns{6}
\tablecaption{Table of Observations, Radio Properties\label{tbl:log}}
\tablehead{\colhead{Source} & \colhead{Obsn Dates} & \colhead{TOS} &  \colhead{array config.}  &\colhead{F$_{R}$} & \colhead{L$_{R}$}
\\
\colhead{}  & \colhead{} & \colhead{(s)} &  \colhead{} 
& \colhead{($\mu$Jy)} &  \colhead{(10$^{14}$ erg s$^{-1}$ Hz$^{-1}$)}
}
\startdata
2M1207 & 2005/01/11--19 &30292  &    BnA &  $<$29 &  $<$0.98 \\
TWA~5B  & 2002/02/20\tablenotemark{a} &850 &  A &$<$120 & $<$2.8\\
       & 2005/06/09 &6150 &  CnB & $<$84    & $<$2.0\\
SSSPM1102 & 2005/10/29 &12350  & DnC & $<$42  & $<$1.6\\
\enddata
\tablenotetext{a}{Archival observation from project AC605}
\end{deluxetable}

\end{document}